\documentclass[
 reprint,
superscriptaddress,
 amsmath,amssymb,
 aps,
prb,
]{revtex4-1}

	\usepackage{bm}
	\usepackage{braket}
	\usepackage{dsfont}

 	\usepackage{graphicx}
	\usepackage[caption=false]{subfig}
	\usepackage[export]{adjustbox}

	\newcommand{\vect}[1]{\boldsymbol{#1}}		
	\newcommand{\op}[1]{\hat{\boldsymbol{#1}}}	
	\newcommand{\hbn}{{\it h}BN}


\begin{document}

\title{Criteria for deterministic single-photon emission in two-dimensional atomic crystals}

\author{Joshua J. P. Thompson}
\email{thompson@chalmers.se}
\affiliation{ Department of Physics, Chalmers University of Technology, 412 96 Gothenburg, Sweden}
\author{Samuel Brem}
\affiliation{ Department of Physics, Chalmers University of Technology, 412 96 Gothenburg, Sweden}
\author{Hanlin Fang}
\affiliation{Department of Microtechnology and Nanoscience (MC2), Chalmers University of Technology, 412 96 Gothenburg, Sweden}
\author{Joey Frey}
\affiliation{Department of Microtechnology and Nanoscience (MC2), Chalmers University of Technology, 412 96 Gothenburg, Sweden}
\author{Saroj P. Dash}
\affiliation{Department of Microtechnology and Nanoscience (MC2), Chalmers University of Technology, 412 96 Gothenburg, Sweden}
\author{Witlef Wieczorek}
\affiliation{Department of Microtechnology and Nanoscience (MC2), Chalmers University of Technology, 412 96 Gothenburg, Sweden}
\author{Ermin Malic}
\affiliation{ Department of Physics, Chalmers University of Technology, 412 96 Gothenburg, Sweden}

\date{\today}

\begin{abstract}
The deterministic production of single-photons from two dimensional materials promises to usher in a new generation of photonic quantum devices. In this work, we outline criteria by which single-photon emission can be realised in two dimensional materials: \textit{spatial isolation, spectral filtering} and \textit{low excitation} of quantum emitters. We explore how these criteria can be fulfilled in atomically thin transition metal dichalcogenides, where excitonic physics dictates the observed photoemission. In particular, we model the effect of defects and localised strain, in accordance with the most common experimental realisations, on the photon statistics of emitted light. Moreover, we demonstrate that an optical cavity has a negative impact on the photon statistics, suppressing the single-photon character of the emission by diminishing the effect of spectral filtering on the emitted light. Our work provides a theoretical framework revealing criteria necessary to facilitate single-photon emission in two-dimensional materials and thus can guide future experimental studies in this field. 
\end{abstract}

\maketitle

\section{Introduction}
 Designing efficient and reliable single-photon sources will be fundamental in the realisation of scalable quantum communications devices \cite{o2009photonic, aharonovich2016solid}. A single-photon source is a light source, which emits one photon at a time into an individual photon mode.
 Since the first observation of single photons in an atomic system \cite{clauser1974experimental}, the phenomena has been observed in parametric down conversion \cite{burnham1970observation}, quantum dots \cite{michler2000quantum}, cold atoms \cite{farrera2016generation} and crystal defect states \cite{kurtsiefer2000stable}. More recently, two dimensional atomic crystals, such as \hbn \cite{tran2016quantum, chejanovsky2016structural, grosso2017tunable} and atomically thin transition metal dichalcogenides (TMDs) \cite{srivastava2015optically, he2015single, chakraborty2015voltage, tonndorf2015single, kern2016nanoscale, kumar2015strain, palacios2017large, branny2017deterministic}, have emerged as a novel platform in which to realise single-photon emission (SPE). The truly two dimensional nature of these 2D materials enables large quantum efficiencies, offers a straight-foward integration into existing photonic chip technology and offers a rich playground, in which to manipulate their electronic properties. Strain \cite{castellanos2013local, niehues2018strain, feierabend2017impact}, defects \cite{jiang2019defect} and patterned dielectrics \cite{shi2019gate} affect electronic and optical properties much more significantly than in bulk materials, while heterobilayers \cite{li2016heterostructures, nagler2019interlayer}, moir\'{e} physics \cite{yu2017moire, lu2019modulated}, proximity-induced effects \cite{scharf2017magnetic} and Janus monolayers \cite{lu2017janus} offer promising possibilities to  tailor device properties.
 
 In order to observe single-photon emission in two dimensional materials, it is necessary to create local modifications in the electronic properties, both to differentiate from the photons emanating from the unmodified material, but also to limit the number of charge carriers involved in the photoemission process. In \hbn, atomic defects in the material form trapping potentials leading to single-photon emission \cite{ chejanovsky2016structural}. These trapping potentials can also be realised using local strain in TMDs \cite{ kumar2015strain, palacios2017large, branny2017deterministic}. Furthermore it has been shown that strain can tune and activate the photoemission from defect states \cite{grosso2017tunable}.

\begin{figure} [t!]
\includegraphics[clip,width=0.99\columnwidth,valign=t]{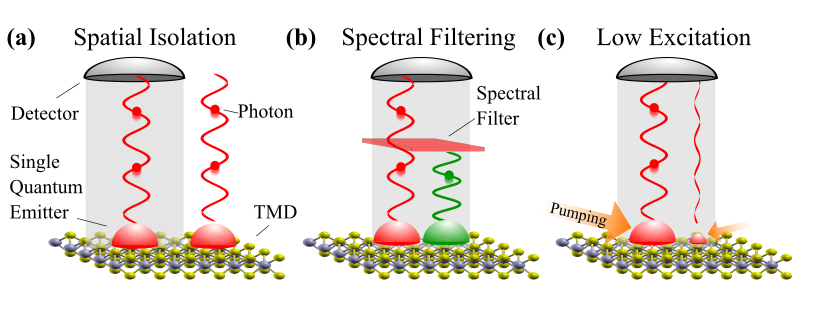}\\
\caption{Schematic illustrating the general criteria for single-photon emission. (a) Spatial Isolation: Single quantum emitters (SQEs) are spaced far apart such that only photons  emitted from one emitter are captured by the detector. (b) Spectral Filtering:  SQEs exhibit different emission frequencies, so that photons from all but one can be obstructed by a filter. (c) Low Excitation: SQEs emitting at the same frequency are excited via optical or electric pumping at different rates, such that the emission from one SQE is most efficient.}
\label{fig:DeviceAndDOSMiniBand}
\end{figure}
	
In this work, we discuss criteria for generating single-photon emission in 2D materials and how these can be realised in realistic experiments exploiting defects and local strain induced by nanopillars.
In order to generate SPE, it is necessary to isolate single quantum emitters. This can happen via spatial isolation, spectral filtering or low-excitation pumping of an ensemble of emitters within a 2D material, cf. Fig. 1.  

The first criterion,  spatial isolation, can be reached by ensuring that the separation between individual emitters is larger than the detection spot size (Fig. 1a). While the sheer number of quantum emitters in a typical 2D material is large,  by ensuring a low defect density \cite{grosso2017tunable}, or in the case of local strain, ensuring that nanopillars are sufficiently spaced \cite{branny2017deterministic, palacios2017large}, SPE can be observed.

The second criterion is spectral filtering (Fig. 1b). In typical SPE experiments, some form of spectral filtering is necessary to filter out all photons except for those within a desired frequency range.  Even in a system with a high emitter density, such that the first criterion is broken, sufficiently detuning their emission frequencies allows only the photons from few quantum emitters to be detected. This detuning can appear naturally \cite{kern2016nanoscale}, particularly when the single-photon emitters have a defect origin. Otherwise it is has been shown that  strain \cite{branny2017deterministic, palacios2017large},  and moir\'{e} potential \cite{yu2017moire, baek2020highly} can lead to sufficiently detuned quantum emitters in 2D materials. Finally, the third criterion is the low excitation of quantum emitters (Fig. 1c). Assuming that the excitation strength driving the occupation of single quantum emitter states is low enough, it is possible to ensure that multiple quantum emitters are occupied differently. This can arise naturally as a result of charge carriers finding the lowest possible energy state, or by device engineering such as strain funnelling \cite{mangu2017strain}.   

In this work, we describe a theoretical framework to calculate the photon statistics in 2D materials. Inspired by common experimental realisations, we model the effect of local strain and defects, and demonstrate how these can satisfy the criteria for single photon emission illustrated in Fig. 1. 

\section{Theoretical Model}

Single-photon emission is typically measured in a Hanbury Brown Twiss (HBT) setup \cite{brown1956test} and is quantified by the second-order autocorrelation function $g_2(t,\tau)$. Here, $\tau$ represents the time delay between the measurements $I_1(t)$  and  $I_2(t+\tau)$. Pure SPE is characterised by $g_2(t, 0)=0$, while $g_2(t,0)=1$ represents the limit of coherent light. For a single-photon mode $\op{c}(t)$, such as in the case of a cavity, the $g_2(t, \tau)$ function reads\cite{scully1999quantum}
\begin{align}
    g_2(t, \tau)= \dfrac{\langle \op{c}^\dagger(t) \op{c}^\dagger(t+\tau) \op{c}(t+\tau) \op{c}(t) \rangle}{\langle\op{c}^\dagger(t) \op{c}(t)\rangle \langle\op{c}^\dagger(t+\tau) \op{c}(t+\tau)\rangle} .
\end{align}
This formalism relies on a discrete, small number of photon modes. In an experimental setting without a cavity, however, there exists a near continuum of photon modes, into which a single quantum source can emit. The continuum of photon modes $\op{c}_{\vect{k}}$ can be expressed as components of the emitted electric field $\vect{E}^\pm(t,\vect{r})$\cite{scully1999quantum}
\begin{align} \label{efield}
\vect{E}^\pm(t,\vect{r}) =  \sum_{\vect{k},\lambda}  \sqrt{\dfrac{\hbar v_{\vect{k}}}{2 \epsilon_0 V}}\vect{\varepsilon}^\lambda_{\vect{k}}  \op{c}^{(\dagger)}_{\vect{k}} (t) e^{\pm i \vect{k}\cdot \vect{r}},
\end{align}
where $\varepsilon_{\vect{k}}^\lambda$ is the polarisation of the photon, $V$ the quantisation volume,  and $v_{\vect{k}}$ the photon frequency.
A crucial element of our analysis will be the role of spectral filtering (Fig. 1b).  This involves only measuring photons within a certain, small energy  range. To incorporate this into our model,  we restrict the sum in Eq. (\ref{efield}) to photon modes within a small energy window $\sigma$, determined by the experimental setup. Typically this will be of the order of $\sigma \sim \pm 100 \mu$eV for photon energies around $\sim 1-2$ eV.  We can relate the electric field components to the excitonic operators $\op{P}$ ($\op{P}^\dagger)$ using Heisenberg's equation of motion \cite{scully1999quantum}
\begin{align} \label{efield2}
   \vect{E}^\pm(t,\vect{r}) = E_0(\vect{r}) F^{(*)}(\sigma, \gamma, \Delta\omega) \op{P}^{(\dagger)}\left(t-\frac{\vect{r}}{c}\right),
\end{align}
where $\Delta \omega = \omega - \omega_m$ is the energy difference between the resonant exciton energy, $\omega$, and the measured photon energy, $\omega_m$, and $\gamma$  the photon linewidth. Here, $E_0(\vect{r})$ relates the electric field operator to the exciton operator at some retarded time. We obtain a filter function 
$ F(\sigma, \gamma, \Delta \omega) = \tan^{-1} \left(\frac{\Delta \omega+\sigma}{\gamma} \right) - \tan^{-1} \left(\frac{\Delta \omega-\sigma}{\gamma} \right) + \dfrac{i}{2} \left[\log \left(1+\frac{(\Delta \omega+\sigma)^2}{\gamma^2}\right) -\log \left(1+\frac{(\Delta \omega-\sigma)^2}{\gamma^2}\right) \right]$, which has  a Gaussian-like profile for the input parameters outlined in this work.
The $g_2$ function with zero time delay ($\tau=0$) can be expressed in terms of these electric field components \cite{scully1999quantum}
\begin{align} \label{g2}
    g_2(t, \tau) = \dfrac{\langle \vect{E}^-(t,\vect{r})\vect{E}^-(t,\vect{r}) \vect{E}^+(t,\vect{r}) \vect{E}^+(t,\vect{r}) \rangle}{ |\langle \vect{E}^-(t,\vect{r}) \vect{E}^+(t,\vect{r}) \rangle|^2},
\end{align}
which can be described in terms of exciton operators using Eq. (\ref{efield2}).  A cluster expansion approach is used to rewrite Eq. (\ref{g2}) in terms of the exciton occupation ${n_i} = \langle \op{P}_i^\dagger \op{P}_i\rangle$ of different quantum emitters resulting in
\begin{align}
    g_2(t, \tau) = \dfrac{\sum_{k,l} {n}_k(t){n}_l(t+\tau) +\text{I}_2(t,\tau)}{  \left[\sum_k {n}_k(t)+\text{I}_1(t)\right] \left[\sum_l {n}_l(t+\tau)+\text{I}_1(t+\tau)\right] }
\end{align}
where $\text{I}_1$ and $\text{I}_2$ describe the one- and two-photon interference terms \cite{bai2017hanbury}, respectively. We omit these interferences in our calculation as they are only driven when the quantum emitter states fall within the optical excitation spot.

We assume that, following an optical excitation of sufficient energy in the  2D material, the generated excitons become trapped in these quantum emitter states before decaying radiatively - a process described in previous experimental and theoretical works \cite{branny2017deterministic, palacios2017large, feierabend2019optical}. The rate $S_\text{in}$ ($S_\text{out}$), at which excitons enter (leave) the quantum emitter state  is determined by the nature of the quantum emitters as well as the excitation pump fluence. Here, we use previously calculated values in Ref. \onlinecite{feierabend2019optical} with $S_\text{in}=1 \text{ps}^{-1}$,  $S_\text{out} = 0.01 S_\text{in} $. The resulting Boltzmann scattering equation governing the exciton population reads
\begin{align}
 \partial_t{n_i} =-\gamma {n_i} + S_\text{in}(1-{n_i}) - S_\text{out}{n_i}.
\end{align}
 The radiative decay rate $\gamma \approx 1$ meV   can be extracted from experimental and theoretical data, \cite{feierabend2019optical, he2016phonon} where previous experiments on excitons trapped in defect/strain states on TMDs show very narrow, sub-nm linewidths \cite{chakraborty2015voltage, tonndorf2015single, kern2016nanoscale, kumar2015strain, palacios2017large, branny2017deterministic, he2016phonon}. In this work,  we allow the system to reach a steady state, before calculating the emission characteristics and the $g_2$-function.   The exact value that exciton density reaches in the steady-state does not alter the photon statistics and only when the relative exciton densities differ  will the value of $g_2(t, \tau)$ be impacted. 
 
In this work, we also investigate the system of a TMD located within an optical cavity. We model the cavity using a photon probability approach \cite{richter2009few}, now taking into account the re-absorption of emitted photon by the quantum emitter states. We derive a set of differential equations for the probability of finding $n$-photons in the cavity mode $p_n = \langle\ket{n}\bra{n}\rangle$, the photon-assisted polarisation of exciton  $S_j^{n+1}= \left\langle \op{P}_j \ket{n+1}\bra{n} \right\rangle$  and the photon-assisted exciton density $f^n_{j}= \left\langle \op{P}^\dagger_j \op{P}_j\ket{n}\bra{n} \right\rangle$
\begin{align}
   \nonumber \partial_t p_n &= -2 \sum_m \biggl( \sqrt{n}\text{Im} \left[g^*_m S^n_m \right] - \sqrt{n+1}\text{Im} \left[g^*_m S_m^{n+1} \right] \biggr) \\
    &-2\kappa n  p_n  +2\kappa (n+1)  p_{n+1},  \\ \nonumber
    \\
        \partial_t S_j^{n+1} &= -i(E_j - \omega_c-i\gamma) S_j^{n+1} + ig_j\sqrt{n+1} \times\\ \nonumber
    &\times\biggl(f_{j}^{n+1} \bigl( 1-\frac{1}{p_{n+1}}f_{j}^{n+1}  \bigr )-f^n_{j} \bigl( 1-\frac{1}{p_{n}} f^n_{j}\bigr) \biggr ) \\\nonumber
    &-\kappa (2n+1) S_j^{n+1} + \kappa \sqrt{(n+1)(n+2)} S_j^{n+2}, \\\nonumber
    \\
    \partial_t f^n_{j} &=  - 2\sqrt{n+1} \text{Im}\left[g^*_j S^{n+1}_j\right]  -2\kappa n f^n_{j}  \\&+2\kappa \sqrt{n+1} f^{n+1}_{j}+S^e_\text{in}(p_n - f^n_{j})-S^e_\text{out}f^n_{j}.\nonumber
\end{align}
We define the cavity loss as  $\kappa = \omega_C/2Q$, where $\omega_C$ is the cavity mode frequency and $Q$ is the cavity quality factor \cite{vahala2003optical, brem2017microscopic}. The cavity loss, the radiative decay rate $\gamma$ and the exciton-cavity mode coupling $g_j$ determine how quickly the system reaches a steady state.   All photon statistics calculations are taken after the system has reached a steady state, such that the effect of these parameters on $g_2(0)$ can be neglected. 
Within this formalism, the $g_2(0)$ function has the simple form
\begin{align}
    g_2(0) = \dfrac{ \sum_n n(n-1)p_n}{\left(\sum_n n p_n \right)^2}.
\end{align}
Unless otherwise specified, we investigate the exemplary material of monolayer tungsten diselenide (WSe$_2$). However, our model is easily generalisable to other 2D materials.

\section{Spatial Isolation}
The most trivial way to observe SPE is to ensure that, within a given detection spot size, only one quantum emitter exists. In systems, such as artificial quantum dots, this principle is readily employed \cite{michler2000quantum}. In TMDs and other 2D material-based systems, however, the large number of charge carriers renders this difficult. Defects arising during the fabrication process in TMDs are likely to be far too prevalent. Commercially available TMDs show estimated defect densities  of the order of \cite{ mcdonnell2014defect, edelberg2018hundredfold} $10^{13}$ cm$^{-2}$, which is far too high to ensure only one defect per excitation spot size. Quantum emitters arising from localised strain partially address this problem, however it is expected that these local strains lead to a number of possible exciton energy states at the site of the strain \cite{feierabend2019optical} meaning that  both spectral filtering and low-excitation regime are needed to ensure SPE.
Ignoring spectral filtering and low-excitation, the $g_2$ function for degenerate, equally occupied and equally distributed defects states is
\begin{align}\label{defectssimp}
    g_2(t,0) = 1-\dfrac{1}{n_DA_\text{spot}},
\end{align}
where $n_D$ is the defect/emitter density and $A_\text{spot}$ the detection spot area. This predicts that if the number of defects in the detection area $n_DA_\text{spot}$ is larger than 2, SPE (i.e. $g_2<0.5$) will not be observed. We will show in the next section how detuning and spectral filtering can recover single-photon emission in this system.

\begin{figure}[t!]
    \centering
    \includegraphics[width=0.9\linewidth]{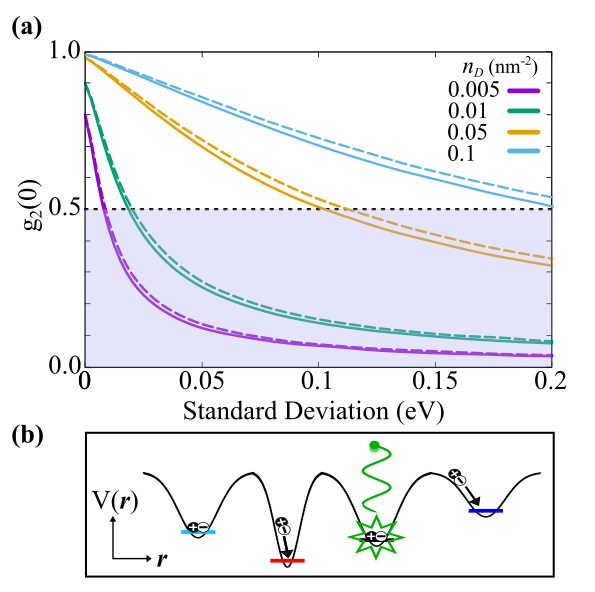}
    \caption{ (a) The second-order autocorrelation function $g_2(0)$ is shown in dependence on the standard deviation of randomly detuned localized excitonic states for several defect densities $n_D$.  The solid (dashed) lines correspond to a filtered energy window of  $\pm$0.2 meV ($\pm$0.5 meV). The detection spot size is fixed at 1000 nm$^2$. The blue area corresponds to the region, in which single-photon emission is observed, i.e. $g_2(0)<0.5$. (b) Schematic showing potential energy landscape due to defects. Excitons, trapped in these defect wells, emit a photon, with wavelength corresponding to depth of the defect potential. }
    \label{fig:my_label}
\end{figure}

\section{Spectral Filtering}
The ability to energetically resolve the photoemission from a material is a fundamental part of  spectroscopy techniques. In this section we describe how the interplay between spectral filtering and detuning of quantum emitters in a 2D material facilitates SPE.

\subsection{Randomly Detuned Defects}
From Eq. (\ref{defectssimp}), the $g_2$ function at zero time delay can be calculated for a number of degenerate defects. We propose that in an experimental device these defects will be detuned and that this could  generate single-photon emission even for large numbers of defects.
Most commonly, defects in TMDs and other two dimensional materials occur during the device fabrication process.  Atomic vacancies, contaminants, substrate surface roughness and charge pockets on the substrate \cite{lin2016defect, jiang2019defect} are just some examples of commonly observed defects. We model this complexity by assuming that, due to the diverse origin of defects in these 2D materials, the resulting quantum emitter states are detuned. The result is shown schematically in Fig. 2 (b), where free excitons become trapped within these detuned defect-based potential wells. The varying depth of these wells determines the emitted photon frequency.

We model this detuning using a Gaussian random distribution with the photon detection window centred on the mean of the distribution. We plot the resulting $g_2$(0) function in Fig. 2 (a) as a function of the standard deviation (the extent of detuning) for a fixed detection spot size  ($A_\text{spot}=$ 1000 nm$^2$) and varying defect density $n_D$. When this detuning is zero we recover the value estimated from Eq. (\ref{defectssimp}). As the detuning increases, the photon energy is more likely to fall outside the detection window, characterised by $F(\sigma, \omega_m,  \gamma, \omega)$ (see the theoretical section) and therefore fewer defects contribute to the measurement of $g_2(t, 0)$. At a high enough detuning the $g_2(t,0)$ function drops to values below 0.5 (indicated by the blue shaded region), demonstrating SPE. The solid and dashed lines show the behaviour for two different spectral filter windows of $\sigma=$ $\pm 200\mu$eV and $\pm 500 \mu$eV, respectively. While the filter with higher spectral window requires slightly higher detuning in order to observe SPE, it is clear that this spectral filtering continues to be applicable for low energy resolutions.

\subsection{Strain-Detuned Quantum Emitters}
It has been recently demonstrated that the application of strain can be used to tune the emission energy of photons from defect sites in \textit{h}BN \cite{grosso2017tunable}. Furthermore, it has been shown that local strain can activate the photoemission from defect states in TMDs, possibly via both detuning and exciton funneling \cite{tonndorf2015single, kern2016nanoscale,  he2016phonon}.
We propose that a local strain gradient leads to position-dependent detuning of the defect sites in the lattice and that this is sufficient to generate SPE. While the exact mechanism of the interplay between these defect states and strain is unknown and would require more complex calculations beyond the scope of this work, we can estimate the effect of a strain gradient on the TMD lattice. This should be generalisable to other types of quantum emitters, such as patterned dielectric or moir\'{e} based emitters, where this strain detuning would be pivotal.

Based on previous  experimental geometries of nanopillars \cite{branny2017deterministic, palacios2017large}, we assume that the TMD rests on a nanopillar and tents slightly. This tenting radius is typically of the order of two times larger than the nanopillar width. Assuming a P\"{o}schl-Teller potential profile $
    V(\vect{r}) = -\frac{\text{V}_\text{max}}{\cosh^2(\alpha \vect{r})}
$  we can estimate the spatial dependence of the strain-induced detuning based solely on the TMD tenting radius and height. The estimated strain potential is in good qualitative agreement with  previous studies \cite{brooks2018theory, feierabend2019optical, chirolli2019strain}.  Here, $\text{V}_\text{max}$ is the potential at the bottom of the confining potential and $\alpha$
is chosen such that  $|\vect{r}|$ equal to the tenting width corresponds to the half maximum.
We use previous work to estimate the strain percentage at the centre of the nanopillar based on the geometry with the strain given as \cite{castellanos2013local, vella2009macroscopic} 
$
    \text{Strain} = \frac{\pi tH}{(1-\nu^2)W^2},
$
where  $t$ is the TMD thickness, $H$ is the height of the TMD on the nanopillar, $W$ is the tenting width and $\nu$ is the Poisson's ratio with values for TMDs extracted from Ref. \onlinecite{fan2015electronic}.
We can estimate how this relates to a detuning of the exciton resonance in various TMD monolayers \cite{niehues2018strain, aas2018strain}. We find spectral red-shifts in the range of  $\sim 50- 300$ meV, which is in good agreement to experimentally observed values at comparable TMD tenting dimensions \cite{castellanos2013local, branny2017deterministic, palacios2017large, kumar2015strain} .

\begin{figure}[t!]
    \centering
    \includegraphics[width=0.9\linewidth]{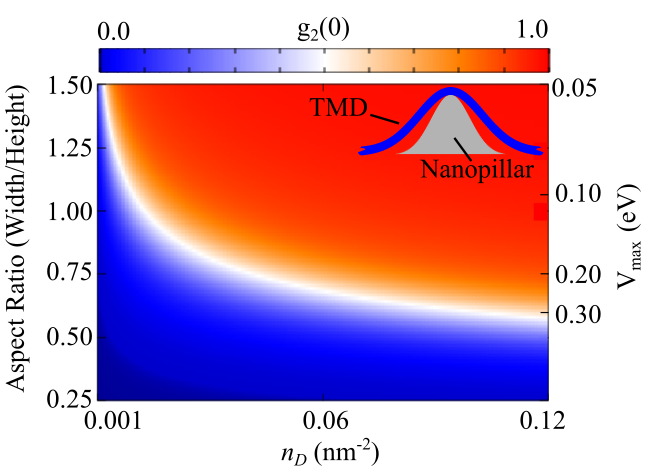}
    \caption{The $g_2(0)$ function for a 2D square array of degenerate single quantum emitters which are detuned by local strain. We present  $g_2(0)$  as a function of the TMD tenting profile aspect ratio (and the corresponding potential maximum) and the single quantum emitter/defect density $n_D$. The white band separates the SPE region (blue) from the non-SPE region (red). The inset is a schematic illustrating a TMD on a nanopillar.}
    \label{fig:my_label}
\end{figure}

The effect of local strain on the $g_2(0)$ function for an array of single quantum emitters is presented in Fig 3.  The height is fixed at 100nm, a typical value  found in experiments \cite{branny2017deterministic, palacios2017large}, and the aspect ratio (width/height) is varied along with the defect density $n_D$.   The $g_2$ measurement is centered around the most detuned photon energy  (corresponding to the centre of the nanopillar).
At high aspect ratio, the area of local strain is much larger than the typical separation between quantum emitters. Therefore, the detuning of the photoemission from these defect sites varies slowly with distance from the centre of the nanopillar resulting in a high $g_2(0)$ function. As the aspect ratio decreases, the detuning potential becomes more narrow in space, detuning fewer quantum emitters, but detuning those more strongly, leading to a $g_2(0)<0.5$ and SPE. As the defect density increases the required aspect ratio needs to be lower in order to facilitate SPE.

\begin{figure}[t!]
    \centering
    \includegraphics[width=0.9\linewidth]{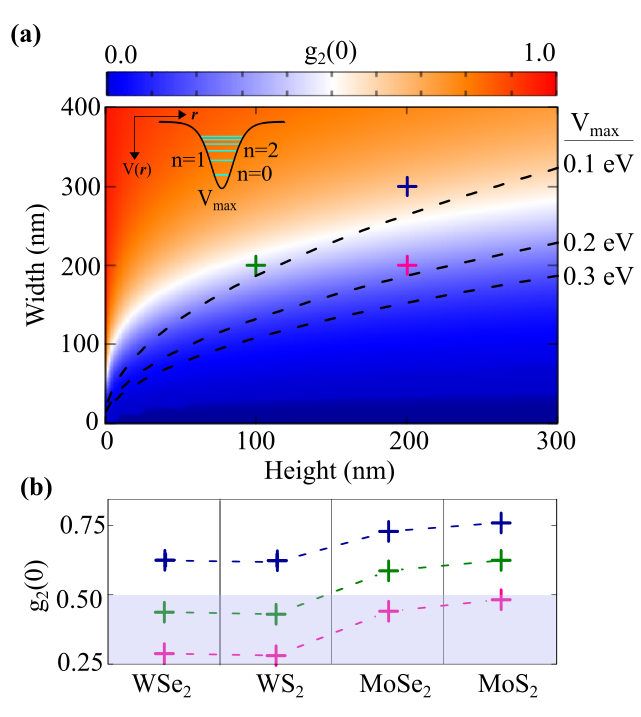}
    \caption{ (a) The $g_2(0)$ function for strain-confined energy levels as a function of the TMD tenting height and width.  The white band separates the SPE region (blue) from the non-SPE region (red). Lines of constant potential well depth $\text{V}_\text{max}$ due to the applied strain are shown with black dashed lines.  The inset illustrates schematically strain-confined energy levels arising from the P\"{o}schl-Teller potential. (b) The $g_2(0)$ function for nanopillar dimensions marked with a cross in (a) evaluated for different TMD materials. The blue area corresponds to the region in which single-photon emission is observed.  }
    \label{fig:my_label}
\end{figure}

\subsection{Strain-induced quantum emitters}

Local strain gradients do not only detune already existing (defect) quantum emitters, they can also create trapping potentials sufficient to localise exciton states \cite{feierabend2019optical, branny2017deterministic, palacios2017large}.  Here, we show that,  for  a given TMD tenting radius and nanopillar height, the $g_2(0)$ function from the resulting quantised energy levels can be calculated. 
As in the previous section, the strain confining potential can be modelled as a P\"{o}schl-Teller potential. The corresponding energy levels are extracted from the two-dimensional Schr\"{o}dinger equation \cite{agboola2010solutions}, assuming an effective exciton  mass of  $M_{\text{MX}_2}=m^*_{c,{\text{MX}_2}}+m^*_{v,{\text{MX}_2}}$, where $m^*_{c/v,{\text{MX}_2}}$ are the electron/hole effective mass of the considered TMD monolayer \cite{kormanyos2015k}. The resulting energy levels, with compound index $n$ are shown in the inset of Fig. 4(a).

We plot $g_2(0)$  as a function of the strained WSe$_2$ profile height and tenting width, cf. Fig 4(a).
While taller nanopillars have a larger potential maximum $V_\text{max}$ (by increasing the strain percentage), narrowing the width increases $V_\text{max}$ and also the separation between consecutive energy levels. This is clear from the dashed lines in Fig 4(a), which show lines of constant $V_\text{max}$.  As a result, narrower and taller tenting profiles are more likely to lead to SPE (blue-shaded region in Fig. 4(a)).

In Fig. 4(b), we show the $g_2(0)$ function for different locally strained TMDs,  with the tenting profile indicated by the crosses in Fig. 4(a). For equivalent levels of strain (i.e. tenting profiles),  tungsten-based TMDs exhibit more pronounced band gap modulation \cite{niehues2018strain} than their molybdenum counterpart and this results in a more pronounced SPE due to a more efficient spectral filtering. While this is one reason why WSe$_2$ is the TMD of choice in SPE experiments, this material also exhibits the most efficient exciton capture rate \cite{chang2013orbital, brooks2018theory} leading to high quantum efficiency. 

\begin{figure}[t!]
    \centering
    \includegraphics[width=0.9\linewidth]{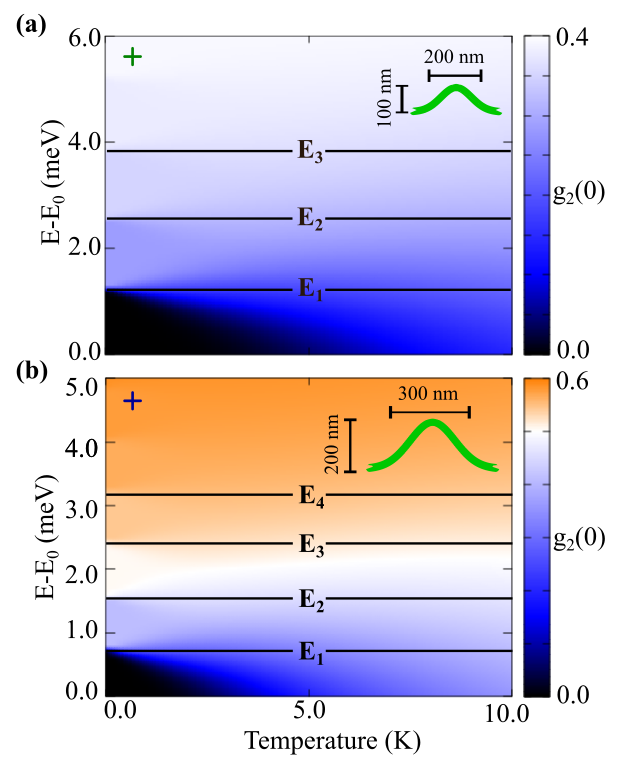}
    \caption{The $g_2(0)$ function in dependence on temperature and Fermi energy. The Fermi level is measured relative to the lowest energy $E_0$. We show plots for two TMD tenting profile dimensions of (a) 200 nm $\times$ 100 nm and (b) 300 nm $\times$ 200 nm (marked with crosses in Fig. 4(a)).   The lowest energy levels above $E_0$ are marked with horizontal lines in each sub figure. }
    \label{fig:my_label}
\end{figure}

\section{Low excitation}
The results presented in the previous section represent the worst possible case in terms of SPE, in which all quantum emitters within the detection spot range are equally excited.  In practice, by ensuring the excitation power is low enough, it is possible to only excite a small number of quantum emitter states \cite{grosso2017tunable, branny2017deterministic, palacios2017large}. Furthermore, at low excitations, processes such as the strain-funnelling of free excitons \cite{castellanos2013local, kumar2015strain} , ensure the occupation of a single quantum emitter to be higher facilitating SPE.

We discuss this principle in the context of strain-induced quantum emitters. At low temperatures, excitons that enter this strain trapping potential will fall into the lowest unoccupied energy level. Only once  the pump power  is increased to above the saturation limit of this energy level,  higher energy levels will become filled.

We demonstrate the importance of excitation strength for  two strain-induced TMD tenting profiles, cf. Fig. 5.  We define a Fermi level on the system of localised excitons, in order to simulate the pump-dependent filling of these localised energy levels.  In the previous sections, we assume the Fermi level to be large, $E>E_{N}$ (where $N$ is the highest energy level), such that all quantum emitters are equally filled. Now, we investigate how the Fermi level will increase with the pump fluence and  move through the strain-confined  energy levels. As a direct consequence,  the $g_2(0)$ function increases, as more emitters are involved.  At 0 K, we see  clear steps in the $g_2(0)$ function, exactly at the position of these higher index energy levels $E_{n>0}$, cf. the black horizontal lines in Fig. 5. At higher temperatures, this effect becomes smoothed due to the Fermi-Dirac statistics of these confined states, which describes the saturation behaviour observed experimentally \cite{grosso2017tunable, branny2017deterministic, he2016phonon}. Furthermore, the temperature induced broadening of the energy levels raises the $g_2(0)$ at low Fermi levels.
 Figure 5 demonstrates that the conditions for SPE prepared by just spectral filtering,  can be further improved by applying low excitations. We can even move from the region with $g_2>0.5$ (orange-shaded area in Fig. 5b) to the SPE limit (blue-shaded area) by reducing the Fermi energy.

\section{Optical Cavities}
By enclosing a quantum emitter in a cavity, the Purcell effect\cite{reeves20182d, vogl2019compact} leads to a strong enhancement in the light-matter coupling. As a result, the quantum efficiency and output intensity from these systems is significantly larger than their cavity-free counterpart. While a higher output of photons is beneficial in practical devices, it is necessary to discuss the effect of a cavity on single-photon emission.
While the SPE criteria spatial isolation and low-excitation limit could be preserved, the effect of spectral filtering becomes negligible. In a cavity-free system, detuned quantum emitters emit light into any of the available photon modes, which in a vacuum will be those resonant to their corresponding exciton energy.  In a cavity, however,  while the light matter coupling is significantly larger, only a single-photon mode is supported. As a result, detuning of quantum emitters from the cavity mode simply affects the conversion rate from excitons to photons. For exciton energies close to or equal to the cavity mode we see high conversion rates, whereas extremely off-resonant excitons decay much less readily and thus do not contribute strongly to the number of photons in the cavity. This leads to a decreased $g_2(0)$ in the cavity system. However, for reasonable levels of detuning, the resulting $g_2(0)$ value is only slightly lowered compared to the effect of spectral filtering, as will be discussed below.

\begin{figure}[t!]
    \centering
    \includegraphics[width=0.9\linewidth]{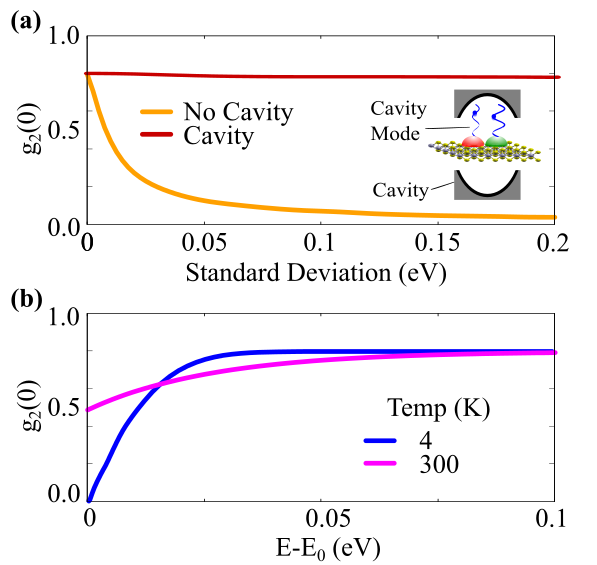}
    \caption{ (a) The $g_2(0)$ function  dependence  on the standard deviation of randomly detuned states comparing   the results with and without a cavity (red and orange, respectively). The quality factor of the cavity is set to $Q=5000$\cite{reeves20182d}, the filtered energy window  is $\sigma =\pm 0.2$ meV and the cavity energy is $1.8$ eV.(b)  $g_2(0)$ as a function of Fermi level for $T=4$K (blue) and $T=300$K (purple) at a fixed standard deviation of $0.01$ eV within an optical cavity. The Fermi level is measured from the lowest defect energy level $E_0$. Schematics of the cavity case is shown in the inset with excitons emitting into the cavity mode. }
    \label{fig:my_label}
\end{figure}

In Fig. 6 (a)  we compare the $g_2(0)$ function for randomly detuned quantum emitters placed within and without a cavity.  We assume that the cavity mode is resonant to the mean detuning energy. 
The cavity-free case (orange) shows a much stronger decrease in the $g_2(0)$ function with the detuning of the states (spectral filtering) than the cavity case (red). The latter barely deviates from the expected non-detuned value $g_2(0)=0.8$ (defined by Eq. (11)) .  By enclosing the system in a cavity, the detuned quantum emitters emit into the cavity mode, rather than the photon mode resonant to the emitter energy. The resulting output photon frequency will be defined by the cavity mode such that spectral filtering cannot be performed. Evidently, we observe that a system contained in a cavity will have a larger $g_2(0)$ function than its equivalent cavity-free setup.

Despite this, either by ensuring one emitter per cavity (spatial isolation) or by  keeping the excitation rate low (low-excitation limit), it is possible that SPE can still be observed in a cavity system \cite{peyskens2019integration, vogl2019compact}.  
In Fig. 6 (b), we demonstrate how low excitation can be used to recover SPE in a cavity system. The $g_2(0)$ function is shown as the  Fermi level is varied  (as in Fig. 5) for the two temperatures of 4K (blue) and 300K (purple). At low temperatures the density of states for a given emitter is centred around a narrow range meaning that tuning the Fermi-level will allow SPE to be recovered. At larger temperatures, the average energy separation between emitter energies is smaller than the thermal broadening. As a result, ensuring emission into the cavity mode from only one quantum emitter becomes more difficult for very small Fermi levels.

\section{Conclusions}
We present a theoretical framework for description of single-photon emission in 2D materials. Three general criteria for single-photon emission are outlined:  spatial isolation, spectral filtering and low-excitation pumping of quantum emitters. 
In particular, we describe how inherent detuning of defect-induced emitters as well as strain modulation can allow single-photon emission. We also discuss how local strain originating from nanopillars can lead to quantised energy levels and that the nanopillar separation can faciliate single-photon emission. We demonstrate that the second-order autocorrelation function $g_2(0)$ depends on the tenting profile of the TMD as it sits on the nanopillar. Furthermore, we show that an optical cavity has a negative impact on single-photon emission, 
since spectral filtering of emitted light becomes washed out. 
 The obtained insights can be generalised to other related materials, such as  hBN and van der Waals heterostructures as well as to materials where single-photon emission is yet to be observed, such as in phosphorene \cite{villegas2016two} and Janus monolayers \cite{lu2017janus}.  

\begin{acknowledgements}
We thank Marten Richter (TU Berlin) for fruitful discussions.
This project has received funding from the Excellence Initiative Nano at Chalmers. Furthermore, we are thankful to the 
Swedish Research Council and the European Unions Horizon 2020 research and innovation programme under grant agreement no. 881603 (Graphene Flagship). W.W. acknowledges financial support by the Carl Tryggers Stiftelse.
\end{acknowledgements}

\bibliographystyle{apsrev4-1}

\end{document}